\documentclass[a4paper,fleqn,usenatbib]{mnras}

\usepackage{amsmath}	
\usepackage{amssymb}	
\usepackage{times,txfonts}

\usepackage[T1]{fontenc}
\usepackage{ae,aecompl}

\usepackage{graphicx}	

\usepackage{wasysym}

\title[Analytic relations among greybody parameters]{Remarkable analytic relations among greybody parameters}

\author[D. Elia et al.]{
Davide Elia,$^{1}$\thanks{E-mail: davide.elia@iaps.inaf.it}
S. Pezzuto,$^{1}$
\\
$^{1}$INAF, Istituto di Astrofisica e Planetologia Spaziali, Via Fosso del Cavaliere 100, I-00133, Roma, Italy}

\date{Accepted XXX. Received YYY; in original form ZZZ}

\pubyear{2016}

\begin{document}
\label{firstpage}
\pagerange{\pageref{firstpage}--\pageref{lastpage}}
\maketitle

\begin{abstract}
In this paper we derive and discuss several implications of the analytic form of 
a modified blackbody, also called greybody, which is widely used in Astrophysics, 
and in particular in the study of star formation in the far-infrared/sub-millimeter 
domain. The research in this area has been greatly improved thanks to recent observations
taken with the \textit{Herschel} satellite, so that it became important to clarify the
sense of the greybody approximation, to suggest possible further uses, and to delimit
its intervals of validity.
First, we discuss the position of the greybody peak, making difference between 
the optically thin and thick regimes. Second, we analyze the behavior of bolometric
quantities as a function of the different greybody parameters. The ratio between the 
bolometric luminosity and the mass of a source, the ratio between the so-called 
``sub-millimeter luminosity'' and the bolometric one, and the bolometric 
temperature are observables used to characterize the evolutionary stage of a source, 
and it is of primary importance to have analytic equations describing the 
dependence of such quantities on the greybody parameters. Here we discuss all these 
aspects, providing analytic relations, illustrating particular cases and providing
graphical examples. Some equations reported
here are well-known in Astrophysics, but are often spread over different publications.
Some of them, instead, are brand new and represent a novelty in Astrophysics literature.
Finally we indicate an alternative way to obtain, under some conditions, the greybody 
temperature and dust emissivity directly from an observing spectral energy distribution, 
avoiding a best-fit procedure. 
\end{abstract}




\section{Introduction}
The concept of blackbody is widely used in modern Astrophysics to model a quantity of
phenomena that approach the ideal case of radiation emitted by an object at the 
thermal equilibrium which is a perfect emitter and absorber; the shape of the spectrum 
is completely described in terms of its temperature. Laws describing global 
characteristics of the blackbody, 
as the Wien's or of Stefan-Boltzmann's ones, represent renowned milestones of quantum Physics.
However, as far as the continuum emission of a source departs from a perfect blackbody behavior 
and another analytic expression is invoked in place of the Planck's function to describe
the corresponding spectral energy distribution (hereafter SED), it becomes interesting 
to understand how the well-known relations valid for a blackbody have to change in turn.

In particular, large and cold interstellar dust grains ($D>0.01~\mu$m, $T\lesssim 20$~K) are 
recognized to be poor radiators at long wavelengths ($\lambda \gtrsim 50~\mu$m), therefore 
their emission requires to be modeled by a blackbody law with a modified emissivity smaller 
than 1 (i.e. the value corresponding to the ideal case), and being a decreasing function of 
wavelength \citep[see, e.g.,][and references therein]{gor95}. The typically adopted expressions
for such emissivity are summarized in Section~\ref{gb} of this paper.

Modeling the dust emission with a modified blackbody (hereafter greybody for the sake of 
brevity) has been widely used to obtain the surface density and (if distance is known) 
the total mass along the line of sight for structures in the Milky Way (diffuse 
clouds, filaments, clumps, cores) 
or for entire external galaxies. Two cases are generally possible: $i)$ an observed 
SED is available, built with at least three spectral points, so that a best-fit is 
performed to determine simultaneously both the column density and the average temperature of 
the emitting source \citep[e.g.,][]{and00,olm09}, or $ii)$ only one flux measurement at a single wavelength (typically in the sub-millimeter regime) is available, then an assumption on the 
temperature value has to be made \citep[e.g.,][]{fau04,moo07} to obtain the column density.

Recently, the availability of large amounts of data from large survey programs for the 
study of star formation with the \textit{Herschel} satellite \citep{pil10}, such as 
Hi-GAL \citep{mol10a}, HGBS \citep{and10}, HOBYS \citep{mot10}, and EPoS \citep{rag12} 
made it possible to build the far-infrared/sub-millimeter five band (70, 160, 
250, 350, and 500~$\mu$m) SEDs of the cold dust in the Milky Way 
\citep[e.g.,][]{eli10,kon10,gia12,pez12,eli13} in a crucial range usually containing the emission 
peak of cold dust. In this case the approach described at the point $i$ represents 
the preferable way for estimating the physical parameters of the greybody which 
best approaches the observed SED. 

In this paper, after introducing in Section~\ref{gb} the 
analytic expression of the greybody, in Section~\ref{gbmax} we discuss how, unlike
the case of a blackbody, the position of the greybody emission peak shifts as a 
function of different parameters besides the temperature. Moreover, further obtainable 
quantities as, for example, bolometric luminosity and temperature, are often used to 
characterize star forming clumps \citep[e.g.,][]{eli10,eli13,gia12,str15}. A 
comparison with the values predicted analytically by the greybody model as a 
function of its physical parameters turn out to be interesting in this respect. We 
derive such functional relationships in Sections~\ref{gbbol} and \ref{tbolsec}.
Further considerations on the obtained analytic relations suggest a
method for deriving the greybody temperature and dust emissivity of an SED without
carrying out a best-fit procedure. This is discussed in Section~\ref{sedbetatemp}.
Finally, in Section~\ref{conclusions} we summarize the obtained results.

\section{The equation of a greybody}\label{gb}
The solution of the radiative transfer equation for a medium with optical depth $\tau_{\nu}$
(function of the observed frequency $\nu$) and for a source function constituted by the Planck's
blackbody $B_{\nu}$ at temperature $T$ is
\begin{equation}
I_{\nu} = (1-\mathrm{e}^{-\tau_{\nu}})\,B_{\nu}(T)\label{igb}
\end{equation}
\citep[cf., e.g.,][]{cho10}, where
\begin{equation}
B_{\nu}(T)=\frac{2\,h\,\nu^3}{c^2}\frac{1}{\mathrm{e}^\frac{h\,\nu}{k_B\,T}-1}\,.
\end{equation} 
With $h$, $k_B$, and $c$ we indicate the Planck's and Boltzmann's constants 
and the light speed in vacuum, respectively. Assuming $I_{\nu}$ being uniform over 
the solid angle $\Omega$, the corresponding flux is
\begin{equation}\label{gbthick}
F_{\nu} = \Omega\,(1-\mathrm{e}^{-\tau_{\nu}})\,B_{\nu}(T)\,.
\end{equation}

The empirical behavior of $\tau_{\nu}$ as a function of $\nu$ for large interstellar dust grains 
is generally modeled as a power law \citep{hil83} with exponent $\beta$:
\begin{equation}\label{eqtau}
\tau_{\nu}=\left( \frac{\nu}{\nu_0} \right)^{\beta}\,,
\end{equation}
where the parameter $\nu_0$ is such that $\tau_{\nu_0}=1$.

In the limit of $\nu \ll \nu_0$, the term $(1-\mathrm{e}^{-\tau_{\nu}})$ can be approximated 
as follows
\begin{equation}
\lim_{\frac{\nu}{\nu_0}\rightarrow 0} \left( 1-\mathrm{e}^{-\tau_{\nu}} \right)=\tau_{\nu}=\left( \frac{\nu}{\nu_0} \right)^{\beta}\,.
\end{equation}
For an opportunely large $\nu_0$ it can happen that all the observed frequencies fall 
in the regime in which the greybody turns out to be optically thin\footnote{We 
call this case optically thin, although also the one described by Equation~\ref{gbthick} 
is optically thin at low frequencies. However, the $\nu \ll \nu_0$ condition ensures 
$\tau_{\nu}$ to be $\ll 1$ across the entire frequency 
range taken into account.}. In such case, Equation~(\ref{igb}) becomes
\begin{equation}\label{igbthin}
I_{\nu} \approx \left( \frac{\nu}{\nu_0} \right)^{\beta} B_{\nu}(T)\,.
\end{equation}

Recalling the definition of optical depth, $\tau_{\nu} \equiv \kappa_{\nu} \int\rho\mathrm{d}s$,
where $\kappa_{\nu}$ is the opacity of the medium, $\rho$ is its volume density and $s$
is the spatial integration variable along the line of sight, in the optically thin regime
it becomes
\begin{equation}\label{taugiusta}
\tau_{\nu} \approx\kappa_{\mathrm{ref}}\left(\frac{\nu}{\nu_{\mathrm{ref}}}\right)^\beta\Sigma,
\end{equation}
where $\Sigma$ is the surface, or column, density, and $\kappa_{\mathrm{ref}}$ is the 
opacity estimated at a reference frequency $\nu_{\mathrm{ref}}$. For an optically thin
envelope, $\Sigma=M/A$ where $M$ is the mass and $A$ is the projected area of the source. 
For a source located at distance $d$, $A=\Omega\,d^2$, so $\Omega=M/(\Sigma \,d^2)=
(M\,\kappa_{\mathrm{ref}})/(\tau\,d^2)$, then Equation~\ref{gbthick} becomes
\begin{equation}\label{gbthin}
F_{\nu} = \frac{M\,\kappa_{\mathrm{ref}}}{d^2}\,\left(\frac{\nu}{\nu_{\mathrm{ref}}}\right)^\beta\,B_{\nu}(T)\,.
\end{equation}

The decision whether the optically thin assumption is valid or not depends on the validity 
of the substitution $\tau$ for $(1-\mathrm{e}^{-\tau})$. In turn, this means that the error 
$\left| [\tau-(1-\mathrm{e}^{-\tau})]/[1-\mathrm{e}^{-\tau}]\right|$ should be negligible 
compared to the data uncertainties. This point is almost always overlooked in the literature. 
Notice that if $\tau=0.2$ the error introduced in the mathematical substitution is 10\%, which 
is negligible only if the fluxes have been measured with a much larger uncertainty. When 
$\tau=0.1$ the error is $\sim$5\% and only when $\tau=0.02$ the error becomes of the order 
of 1\%.

Finally, let us remind the reader that so far we expressed all quantities as functions
of $\nu$, but they can be equivalently formulated in terms of the wavelength $\lambda$.
For example, the optical depth can be expressed also as $\tau=(\lambda/\lambda_{0})^{-\beta}$, 
with $\lambda_{0}=c/\nu_0$.
Furthermore, in the literature regarding dust emission in the far infrared/sub-millimeter,
generally one encounters the quantity $B_{\nu}(T)$, measured in Jy/sr, expressed as a
function of $\lambda$ (in $\mu$m), which the reader has to keep in mind before applying
the equations reported in this paper to specific cases.

\section{The maximum of greybody emission}\label{gbmax}

The peak position of $I_\nu$ can be found by differentiating Equation~\ref{igb} with respect
to $\nu$. Nevertheless, we prefer to start from the optically thin case (Equation~\ref{igbthin}),
which is quite simpler, and can be approached in a way similar to the derivation of the Wien's
displacement law for a blackbody. In this latter case, imposing the derivative of the Planck's 
function to be 0 leads to solve numerically the equation 
\citep[see, e.g.,][]{ryb79}
\begin{equation}\label{wienx}
x=3(1-\mathrm{e}^{-x})\,,
\end{equation}
where $x\equiv h\,\nu/k_B\,T$. The solution of this equation is $x\simeq 2.82$, i.e.
$ \nu_\mathrm{b}/T=5.88\times10^{10}$~Hz~K$^{-1}$.

Similarly, imposing the same condition to the expression in Equation~\ref{igbthin},
the equation to be solved becomes
\begin{equation}\label{maxinu}
x=(3+\beta)(1-\mathrm{e}^{-x})\,,
\end{equation}
which for $\beta=0$ corresponds to Equation~\ref{wienx}. The value of $x$ increases with $\beta$: 
for instance $x=3.92$ for $\beta=1$, and $x=4.97$ for $\beta=2$. This means that for any 
$\beta\ge 1$ putting $1-\mathrm{e}^{-x}$ equal to 1 results in an error smaller than 2\%. So 
we can write
\begin{equation}
x\simeq 3+\beta
\end{equation}
or
\begin{equation}\label{numax}
\nu_\mathrm{p}\simeq\frac{k_B\,T\,(3+\beta)}{h}=20.837\,T\,(3+\beta)\,\,\,[\mathrm{GHz}]\,.
\end{equation}

The corresponding wavelength is given by  
\begin{equation}\label{lambdamax1}
\lambda_{\nu_\mathrm{p}}\simeq\frac{hc}{k_B\,T(3+\beta)}=\frac{1.439}{T(3+\beta)}\,\,\mathrm{cm}\,.
\end{equation}
Let us remind the reader that this is the wavelength at which the peak of $I_{\nu}$ is 
encountered, so Equation~\ref{lambdamax1} does not apply to $I_{\lambda}$ (see below). 
In Figure~\ref{gbbeta}, $I_{\nu}$ as a function of $\lambda$ is shown for
different values of $T$ and $\beta$, highlighting how the peak position varies according
to Equation~\ref{maxinu}. 

Similarly to Equation~\ref{wienx}, the peak wavelength of $I_{\lambda}$ has to be 
calculated starting from 
\begin{equation}\label{lambdaderiv}
\frac{\partial}{\partial \lambda}\left[\left(\frac{\lambda}{\lambda_0}\right)^{-\beta} 
B_{\lambda}(T)\right]=0\,,
\end{equation}
which, leads to an equation like 
\begin{equation}\label{maxilambda}
x=(5+\beta)(1-\mathrm{e}^{-x})\,,
\end{equation}
where $x \equiv 
h\,c/\lambda\,k_B\,T$ in this case. From this equation, for $\beta=0$, one can obtain the 
most used formulation of the Wien's displacement law, the one in terms of $I_{\lambda}$ 
and $\lambda_{\mathrm{p}}$. Since Equations~\ref{maxinu} and~\ref{maxilambda} constitute 
an incompatible system, this gives an alternative demonstration of the known result
$\lambda_{\mathrm{p}}\neq \lambda_{\nu_\mathrm{p}}$, namely the wavelengths 
at which $I_\lambda$ and $I_\nu$ peak, respectively, do not coincide. 

\begin{figure}
\includegraphics[width=8.3cm]{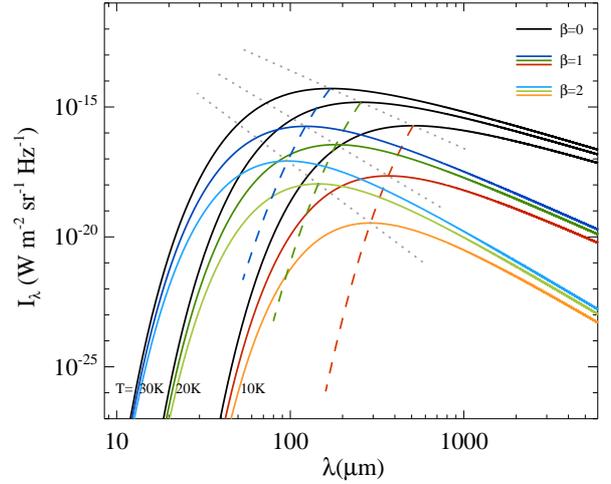}
\caption{Plot of $I_{\nu}$ as a function of $\lambda$ for different values of parameters $T$ 
and $\beta$, in the optically thin case, described by Equation~\ref{igbthin}.
Three temperatures are probed: 10, 20, and 30~K, corresponding to three sets of curves (red, green,
and blue, respectively); for each temperature, three values of $\beta$ are probed: 0 (corresponding 
to the case of a blackbody), 1, and 2, which are plotted, for each set of curves, in black, dark 
color and light color. The frequency $\nu_0$ is chosen such that $\lambda_0=5~\mu$m, in order to 
fulfill the requirements of the optically thin approximation in the considered range of wavelengths.
Dashed lines connect maxima of $I_{\nu}$ at different $\beta$ values, from 0 (top) to 6 (bottom), 
at $T=10$ (red), 20 (green), and 30~K (blue), respectively. Grey dotted lines, instead, connect 
maxima of $I_{\nu}$ at different temperatures, for the cases $\beta=0$ (top), 1 (middle), and 
2 (bottom). All the three lines start at $T=100$~K (top left) down to $T=5$~K (bottom 
right).}\label{gbbeta}
\end{figure}

Let us now consider the most general case. Again, the derivative of $I_{\nu}$ with respect to $\nu$ 
is easier to compute. First of all, let us notice that
\begin{equation}
\frac{\partial (1-\mathrm{e}^{-\tau_{\nu}})}{\partial \nu}=\frac{\beta\tau_{\nu}\mathrm{e}^{-\tau_{\nu}}}{\nu}\,.
\end{equation}
Therefore,
\begin{equation}
\begin{split}
& \frac{\partial}{\partial \nu}\left[(1-\mathrm{e}^{\tau_{\nu}}) B_{\nu}(T)\right]=0\,\Rightarrow\\
& \Rightarrow \frac{2\,h\,\nu^2}{c^2}\frac{1}{\mathrm{e}^{\frac{h\,\nu}{k_B\,T}}-1} \times \\
&\times \left[ \beta\tau_{\nu}\mathrm{e}^{-\tau_{\nu}} +(1-\mathrm{e}^{-\tau_{\nu}})\left( 3-\frac{h\nu}{k_B\,T}\frac{\mathrm{e}^{\frac{h\nu}{k_B\,T}}}{\mathrm{e}^{\frac{h\nu}{k_B\,T}}-1} \right) \right]=0\,.
\end{split}
\end{equation}

Using again $x\equiv h\,\nu/k_B\,T$, the last condition is satisfied if 
\begin{equation}\label{thickpeak}
\frac{\beta\tau_{\nu}}{\mathrm{e}^{\tau_{\nu}}-1}=\frac{x}{1-\mathrm{e}^{-x}}-3\,.
\end{equation}
In the above equation, when $\nu\rightarrow 0$ the two fractions tend to $\beta$ and 1, respectively; 
so the right hand side tends to -2. When $\nu\gg 1$ the left hand side tends to 0, 
while the right hand side tends to $x-3$. Since the left hand side is always positive, the 
solution $\nu_{\mathrm{p}}$ of the equation must be greater than the frequency $\nu_{\mathrm{b}}$ 
at which the right hand becomes positive (notice that $\nu_{\mathrm{b}}$, defined in this way,
coincides with the solution of Equation~\ref{wienx}, which is valid in the case of a pure 
blackbody).

\begin{figure}
\includegraphics[width=8.0cm]{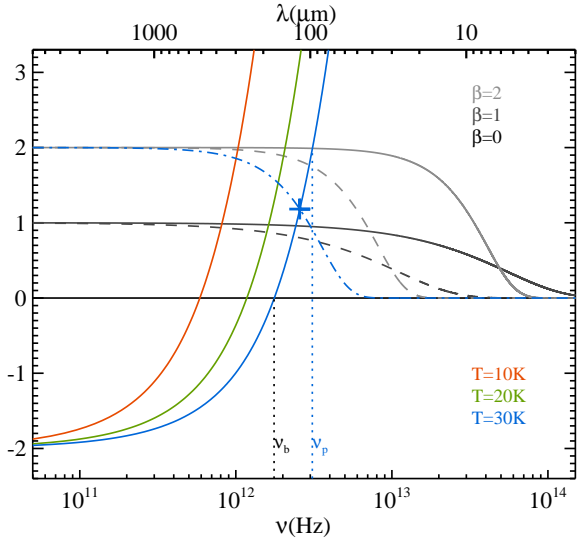}
\caption{The two sides of Equation~\ref{thickpeak} plotted as a function of frequency 
(bottom $x$-axis) and wavelength (top $x$-axis). Units on the $y$-axis are adimensional numbers.
Grey lines represent the left hand side for $\beta=0$ (black), 1 (dark grey), and 2 (light grey), 
respectively, with $\nu_0=30$~THz (solid lines) and $\nu_0=6$~THz (dashed lines), corresponding 
to $\lambda_0=10$ and $50~\mu$m, respectively. In addition, the case $\beta=2$ and 
$\nu_0=\nu_\mathrm{p}$ at $T=30~\mathrm{K}$, namely $\nu_0=2.6$~GHz, (obtained through 
Equation~\ref{nupi}) is plotted as a blue 
dotted-dashed curve. The red, green and blue solid curves represent the right 
hand side for T=10, 20, and 30~K, respectively. For a given choice of the parameters $\beta$, 
$\nu_0$, and $T$, the abscissae of the intersections of the colored curves with the grey ones 
represent the solutions $\nu_\mathrm{p}$ of Equation~\ref{thickpeak}, while the intersections 
with the black curve represent the blackbody case ($\nu_\mathrm{b}$). The fact that
$\nu_\mathrm{p}> \nu_\mathrm{b}$ is highlighted in the case of $T=30$~K through two vertical
dotted lines at the positions of $\nu_\mathrm{b}$ (black) and $\nu_\mathrm{p}$ at $\beta=2$
(blue). The intersection between the two blue lines, i.e. the solid and the dotted-dashed 
one, is marked with a blue cross and is discussed in the text.}\label{soluzione}
\end{figure}

In Figure~\ref{soluzione} the two sides of Equation~\ref{thickpeak} are plotted vs the 
frequency, in correspondence of different choices of the parameters $\beta$, $\nu_0$, and 
$T$. It is noteworthy that the left hand side of the equation depend only on the first 
two of these parameters, while the right hand side depends only on the third one. 
In this figure one can find a graphical representation of the $\nu_\mathrm{p}>
\nu_\mathrm{b}$ condition.

Note that in the limiting case $\nu_0\rightarrow0$ (a greybody optically thick at all frequencies) 
one obtains $\nu_\mathrm{p}\rightarrow\nu_\mathrm{b}$.

The opposite limiting case is $\nu_0\rightarrow\infty$ (a greybody optically thin at all frequencies),
in which the left hand side of Equation~\ref{thickpeak} gets constantly equal to $\beta$ and the frequency 
of the peak corresponds to the solution of Equation~\ref{numax}.

In summary, combining Equations~\ref{numax} and~\ref{thickpeak} one finds
\begin{equation}
1<\nu_\mathrm{p}/\nu_\mathrm{b}\le\left\{
\begin{tabular}{cc}
1.43&$\beta=1$\\
1.79&$\beta=2$\\
2.14&$\beta=3$
\end{tabular}\right.
\end{equation}

So, even though $\nu_0$ can be any real number, nonetheless the peak frequency of the 
greybody lies within a limited range of values that can be parametrized in terms of 
the peak frequency of the blackbody.


We can go further on in extracting information from Equation~\ref{thickpeak}, which
gives the peak of the greybody for any given set of the three parameters $\nu_0$, $T$, and $\beta$. 
However, if the peak is fixed to the frequency $\nu_\mathrm{p}$, then only two parameters 
remain free. If we also impose the condition $\nu_\mathrm{p}=\nu_0$, equivalent to assume 
$\tau_{\nu_\mathrm{p}}=1$, only one parameter is left free, and Equation~\ref{thickpeak} becomes
\begin{equation}\label{thickpeakapprox}
\frac{\beta}{\mathrm{e}-1}=\frac{x}{1-\mathrm{e}^{-x}}-3\,,
\end{equation}
which, for any chosen $\nu_\mathrm{p}$, gives the relation between $T$ and 
$\beta$: if one fixes, say, $T$, then Equation~\ref{thickpeakapprox} gives the value 
of $\beta$ such that $\nu_0=\nu_\mathrm{p}$, and vice versa. We give a graphical
example of this for $T=30$~K and $\beta=2$ in Figure~\ref{soluzione}, highlighting 
$\nu_{\mathrm{p}}=\nu_0$ with a blue cross.

The numerical solutions of Equation~\ref{thickpeakapprox} are $x=3.47,4.10,4.70$ 
for $\beta=1,2,3$, respectively. With these values for $x$, it is possible to put 
$x/\left[1-\mathrm{e^{-x}}\right]\simeq x$. The error decreases from 3\% for 
$\beta=1$ to less than 1.1\% for $\beta=3$. So, for $\tau_{\nu_\mathrm{p}}=1$,
\begin{equation}\label{nupi}
\nu_\mathrm{p}=\frac{k_B\,T}{h}\left(\frac{\beta}{\mathrm{e}-1}+3\right)\,.
\end{equation}
In the above form, Equation~\ref{nupi} gives, for any pair ($T,\beta$), the 
frequency of the peak of a greybody such that $\tau_{\nu_{\mathrm{p}}}=1$; for instance, 
when $T=10$~K  one finds $\nu_0= 746,868,989$~GHz (in terms of wavelengths, 
$\lambda_0=402,346,303~\mu$m) for $\beta=1,2,3$, respectively. For other temperatures, note 
that $\nu_0$ scales linearly with $T$. Now we invert the problem: 
we fix $\lambda_0$ and $\beta$ and look for the values of $T$ that make the peak of the 
greybody falling at $\lambda_\mathrm{p}=\lambda_0$. Dealing, for example, with \textit{Herschel}, 
it is natural to set $\lambda_0=70~\mu$m. Then we find $T=57$~K for $\beta=1$, and $T=43$~K for $\beta=3$. 
So, the triple ($\lambda_0=70~\mu$m, $T=43$~K, $\beta=3$) is such that 
$\lambda_\mathrm{p}=\lambda_0$ and then $\tau(\lambda=70\mu\mathrm{m}) =1$. 
If we keep constant $\lambda_0$ and consider $T \le 43$~K and 
$\beta \le 3$, then the wavelength of the peak shifts to $\lambda_\mathrm{p}>\lambda_0$ (e.g., 
for $\lambda=70$~$\mu$m, $T=40$~K and $\beta=2.5$ from Equation~\ref{thickpeak} we find 
$\lambda_\mathrm{p}\sim~78$~$\mu$m) so that 
$\tau_{\nu_\mathrm{p}}=(\nu_\mathrm{p}/\nu_0)^\beta<1$. In 
conclusion, as long as the temperature of the greybody is less than 43~K 
\citep[the typical case encountered in recent \textit{Herschel} literature, e.g.,][]{gia12,eli13}
we are sure that $\tau_{\nu_\mathrm{p}} \le 1$, independently of the values of 
$T,\beta$ and, for $\lambda_0 \le 70~\mu$m, of $\nu_0$, as long as $\beta \le 3$.

The main limitation of this conclusion is that, actually, it does depend 
on $\nu_0$ which, in general, is not known even if the sources observed with 
Herschel typically have $\lambda_0 \lesssim 70~\mu$m.
In any case, to go further also when only $T$ and $\beta$ are known we proceed as 
follows: first we note that if $\tau_{\nu_{\mathrm{p}}}<1$ then 
$\tau_{\nu_{\mathrm{p}}}/(\mathrm{e}^{\tau_{\nu_{\mathrm{p}}}}-1)>0.582$ 
(this ratio tends to 1 for $\tau_{\nu_{\mathrm{p}}} \rightarrow 0$). This 
condition implies that 
\begin{equation}
\frac{1}{\beta}\left(\frac{x_\mathrm{p}}{1-\mathrm{e}^{-x_\mathrm{p}}}-3\right)>0.582
\Rightarrow\frac{x_\mathrm{p}}{1-\mathrm{e}^{-x_\mathrm{p}}}>\frac{\beta}{\mathrm{e}-1}+3\,.
\end{equation}

For $\beta=1$, $x_\mathrm{p} \gtrsim 3.58$, whilst for $\beta=3$, $x_\mathrm{p} \gtrsim 4.70$; 
with these values of $x_\mathrm{p}$ one can assume $1-\mathrm{e}^{-x_\mathrm{p}} \simeq 1$, 
the error being about 3\% for $\beta=1$ and even lower for higher $\beta$. So, under
the condition $\tau_{\nu_{\mathrm{p}}}<1$, we can cast Equation~\ref{nupi} in the form
\begin{displaymath}
x_{\mathrm{p}}=\frac{h\,c}{k_\mathrm{B}\,\lambda_\mathrm{p}\,T} > \frac{\beta}{\mathrm{e}-1}+3\,,
\end{displaymath}
then, finally,
\begin{equation}\label{lambdap}
\lambda_\mathrm{p} < \frac{h\,c}{k_{\mathrm{B}}\,T}\frac{\mathrm{e}-1}{\beta+3(\mathrm{e}-1)}\,.
\end{equation}

This result can be clearly seen in Figure~\ref{soluzione}, where the blue cross symbol
represents the right hand side of Equation~\ref{lambdap} for $T=30$~K and $\beta=2$.
All family of curves representing the left hand side of Equation~\ref{thickpeak} 
intersecting, in this case ($T=30$~K), the blue solid curve below
the cross symbol have $\tau_{\nu_\mathrm{p}} >1$ and ${\lambda_\mathrm{p}}$
violating the condition imposed in Equation~\ref{lambdap}. Clearly, varying the temperature
would change the position of the cross symbol in that diagram.

If we use Equation~\ref{igbthin} to fit a SED known over a set of fluxes at wavelengths 
longer than a certain $\lambda_{\mathrm{min}}$, to obtain a reliable estimate of $T$
it should be that $\lambda_{\mathrm{p}}>\lambda_{\mathrm{min}}$; if this is the case, 
the condition to have $\tau_{\nu_{\mathrm{p}}}<1$ becomes
\begin{equation}\label{lambdamin}
\lambda_\mathrm{min}<\frac{h\,c}{k_{\mathrm{B}}\,T}\frac{\mathrm{e}-1}{\beta+3(\mathrm{e}-1)}\,.
\end{equation}

Solving for $T$ and setting, for example, $\lambda_\mathrm{min}=70$~$\mu$m we obtain
$T \lesssim 45$~K for $\beta \le 3$; but $T \lesssim 31$~K is found for 
$\lambda_{\mathrm{min}}=100$~$\mu$m and $T \lesssim 20$~K for $\lambda_\mathrm{min}=160$~$\mu$m.

A couple of comments can be made: first, if one uses Equation~\ref{lambdamax1} to 
find the peak of the greybody, then Equation~\ref{lambdamin} is always verified. This 
happens because Equation~\ref{lambdamax1} is valid if $\tau_{\nu_{\mathrm{p}}} \ll 1$ while 
Equation~\ref{lambdamin} is more general, having imposed only that 
$\tau_{\nu_{\mathrm{p}}} < 1$. Second, the condition $\tau_{\nu_{\mathrm{p}}} < 1$ means 
that $\lambda_0 < \lambda_\mathrm{p}$; the assumption $\lambda_{\mathrm{min}}<\lambda_{\mathrm{p}}$ 
does not imply that $\lambda_0<\lambda_\mathrm{min}$, so that Equation~\ref{lambdamin} gives a 
necessary but not sufficient condition to justify the use of Equation~\ref{igbthin}.
In other words, the relation $\lambda_{\mathrm{min}}<\lambda_0<\lambda_\mathrm{p}$ 
is compatible with Equation 23 and in this case it is still true that $\tau_{\nu_{\mathrm{p}}}<1$; 
but then there is a portion of the SED, between $\lambda_\mathrm{min}$ and $\lambda_0$, where 
$\tau_\nu>1$ so that the use of Equation~\ref{igbthin} is not justified over the whole 
observed SED.

We conclude this section by noting that Equation~\ref{lambdamin} is a condition to 
have $\tau_{\nu_{\mathrm{p}}}<1$, not a condition to have a reliable fit of the SED: if the peak 
falls at wavelengths shorter than $\lambda_{\mathrm{min}}$ it is still possible to obtain a 
good fit of the SED, at least if it is not that $\lambda_\mathrm{p} \ll \lambda_\mathrm{min}$. 
However if we use Equation~\ref{igbthin} to fit the SED then, by combining Equations~\ref{maxinu} 
and~\ref{lambdap}, we get the condition $\beta(\mathrm{e}-2)>0$ which is always true: 
this, in turn, means that the condition $\tau_{\nu_{\mathrm{p}}}<1$ is implied by the 
adopted functional form of the SED, and can not be verified a posteriori from the 
values of $T$ and $\beta$ derived from the fit.

\section{Greybody luminosity}\label{gbbol}

The bolometric luminosity, namely the power output of a given source across all wavelengths,
is an observable widely used in several fields of Astrophysics. In particular, in the 
far infrared/sub-millimeter study of early phases of star formation, this quantity is 
exploited in combination with other quantities to infer the evolutionary stage of 
young sources \citep[e.g.,][]{mye98,mol08} as far as their
continuum emission departs from that of a simple cold greybody ($T \sim 10$~K) and 
starts to show signatures of ongoing star formation in form of emission excess at 
shorter wavelengths \citep[$\lambda \lesssim 70~\mu$m, e.g.,][]{eli13}.  

For making a comparison with the luminosity of a simple greybody, analytic dependence
of it on $T$ and $\beta$ has to be explored.

First of all, let us recall the Stefan-Boltzmann's law for a 
black body, describing the power $W_{\mathrm{b}}$ radiated from a black body (per unit 
surface area), calculated as the integral over half-sphere\footnote{For a generic 
solid angle $\Omega$, in Equation~\ref{sbeq} one can replace 
$\pi$ with $\Omega$.}, as a function of its temperature:
\begin{equation}\label{sbeq}
W_{\mathrm{b}}=\pi \int_0^\infty B_{\nu}(T)\,\mathrm{d}\nu=\sigma T^4~,
\end{equation}
where $\sigma=5.67\times 10^{-8}$~W~m$^{-2}$~K$^{-4}$ is the Stefan-Boltzmann constant. 

Here we search for an analogous relation for a generic greybody with exponent $\beta$, in 
the optically thin case (Equation~\ref{igbthin}) :

\begin{equation}
\begin{split}
&W_{\mathrm{g}}&=&\frac{\pi}{\nu_0^\beta} \int_0^\infty {\nu}^{\beta}\,B_{\nu}(T)\,\mathrm{d}\nu=&\\
&&=&\frac{\pi}{\nu_0^\beta} 
\int_0^\infty\frac{2\,h\,\nu^{3+\beta}}{c^2}\frac{1}{\mathrm{e}^{\frac{h\,\nu}{k_B\,T}}-1}\,\mathrm{d}\nu=&\\
&&=&\frac{2\,\pi\,k_B^{3+\beta}\,T^{3+\beta}}{h^{2+\beta}\,c^2\,\nu_0^\beta}
\int_0^\infty\left( \frac{h\, \nu}{k_B\, T}\right)^{3+\beta}
\frac{1}{\mathrm{e}^{\frac{h\,\nu}{k_B\,T}}-1}\,\mathrm{d}\nu\;.
\end{split}
\end{equation}
Imposing $x\equiv\frac{h\nu}{k_B T}$, then $\mathrm{d}x=\frac{h}{k_B T}\mathrm{d}\nu$,
\begin{equation}
W_{\mathrm{g}}=\frac{2\, \pi\, k_B^{4+\beta}\, T^{4+\beta}}{h^{3+\beta}\,c^2\,\nu_0^\beta}
\int_0^\infty\frac{x^{3+\beta}}{\mathrm{e}^x-1}\mathrm{d}x\,,
\end{equation}
which shows a similar power-law dependence on temperature as in Equation~\ref{sbeq}, 
being $\int_0^\infty 
x^{3+\beta}/\left(\mathrm{e}^x-1\right)~\mathrm{d}x$ not depending on temperature. Focusing the attention on this integral, let us notice that $1/(\mathrm{e}^x-1)=\mathrm{e}^{-x}/(1-\mathrm{e}^{-x})
=\sum_{n=1}^\infty \mathrm{e}^{-nx}$, then
\begin{equation}
\int_0^\infty\frac{x^{3+\beta}}{\mathrm{e}^x-1}\,\mathrm{d}x=
\sum_{n=1}^\infty \int_0^\infty x^{3+\beta}\,\mathrm{e}^{-nx}\,\mathrm{d}x\,
\end{equation}
where $\int_0^\infty x^{3+\beta}\,\mathrm{e}^{-nx}\,\mathrm{d}x$ can be integrated by parts recursively. However,
recalling the definition of the Euler's gamma function $\Gamma(z)\equiv 
\int_0^\infty x^{z-1}\,\mathrm{e}^{-x}\,\mathrm{d}z$, and imposing $y = n x$ (then $\mathrm{d}x= 1/n\,dy$), one finds 
\begin{equation}
\begin{split}
&\int_0^\infty x^{3+\beta}\,\mathrm{e}^{-nx}\,\mathrm{d}x=\frac{1}{n^{4+\beta}}\int_0^\infty y^{3+\beta}\,\mathrm{e}^{-y}\,\mathrm{d}y=\\
&=\frac{1}{n^{4+\beta}}\Gamma(4+\beta)\,.
\end{split}
\end{equation}
Again, recalling the definition of Riemann's zeta function $\zeta(z)=\sum_{n=1}^\infty 1/n^z$, 
one finally finds 
\begin{equation}\label{gammazeta}
\int_0^\infty\frac{x^{3+\beta}}{\mathrm{e}^x-1}\mathrm{d}x=\zeta(4+\beta)\Gamma(4+\beta)\,.
\end{equation}
Therefore,
\begin{equation}\label{intgb}
W_{\mathrm{g}}=\frac{2\,\pi\, k_B^{4+\beta}\, \zeta(4+\beta)\,\Gamma(4+\beta)}
{h^{3+\beta}\,c^2\,\nu_0^\beta}~T^{4+\beta}\,.
\end{equation}

Reminding the reader that, for an integer argument $n$, $\Gamma(n)=(n-1)!$, and that the 
$\zeta$ function can be analytically computed for positive even integer arguments, in 
Table~\ref{gz} we quote few representative values of $\Gamma(4+\beta)\zeta(4+\beta)$.

\begin{table}
\centering
\caption{The product $\Gamma(4+\beta)/\zeta(4+\beta)$ for a few values of $\beta$.}
\label{gz}
\begin{tabular}{ccccc}
\hline
$\beta$ & $4+\beta$ & $\Gamma(4+\beta)$ & $\zeta(4+\beta)$ & $\Gamma(4+\beta)\zeta(4+\beta)$ \\
\hline
  0 &   4 &    6 & $\pi^4/90$ & $\pi^4/15$  \\
 0.5 &  4.5 &    11.63 &  1.055 &   12.27  \\
  1 &   5 &   24 &  1.037 &   24.89  \\
 1.5 &  5.5 &    52.34 &  1.025 &   53.66  \\
  2 &   6 &  120 & $\pi^6/945$ & $8\pi^6/63$  \\
 2.5 &  6.5 &   287.89 &  1.012 &  291.34  \\
  3 &   7 &  720 &  1.008 &  726.01  \\
\hline
\end{tabular}
\end{table}

\begin{figure}
\includegraphics[width=8.0cm]{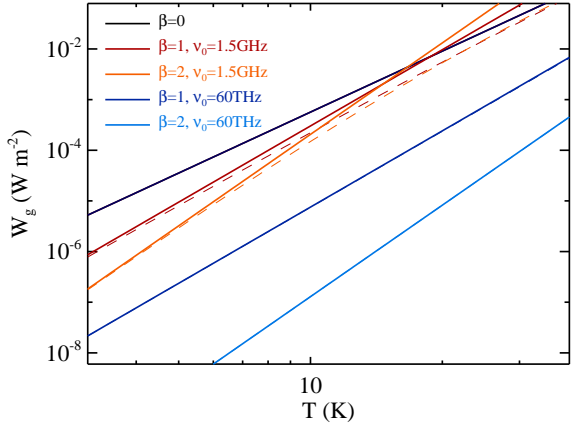}
\caption{Power radiated by an optically thin greybody vs temperature, for different 
choices of $\beta$ and $\nu_0$. The black solid line corresponds to the blackbody case 
($\beta=0$), while red solid lines correspond to $\nu_0=1.5$~GHz (i.e. $\lambda_0=200~\mu$m; 
dark red: $\beta=1$; light red: $\beta=2$) and blue solid lines correspond to $\nu_0=60$~THz 
(i.e. $\lambda_0=5~\mu$m; dark blue: $\beta=1$; light blue: $\beta=2$). Dashed lines 
represent the optically thick greybody (Equation~\ref{igb}): the same color corresponds to 
the same parameter combination reported above. Note that for $\nu_0=60$~THz the two 
regimes turn out to be indistinguishable in the temperature range shown here, so that 
dashed lines are completely superposed on the solid lines, then invisible
The point where the power of the greybody intersects the power of the 
blackbody marks the temperature above which the condition of optically thin medium 
is violated and Equation~\ref{intgb} is no longer valid.}\label{lbols}
\end{figure}

In Figure~\ref{lbols} the $W_{\mathrm{g}}$ vs $T$ relation is displayed for some 
choices of the parameters $\beta$ (including the blackbody case) and $\nu_0$. 
Also in this case it is possible to notice that the blackbody is the best radiator
at the probed temperatures, but all lines with $\beta > 0$ appear steeper than the 
blackbody one, therefore there is an intersection point at some $T_\mathrm{int}$
such that a given line is higher than the blackbody one at $T > T_\mathrm{int}$.
This situation is unphysical because no thermal spectrum can radiate more than 
the blackbody at the same temperature. This means that above $T_\mathrm{int}$ 
the hypothesis of optically thin medium is violated and to 
use Equation~\ref{igbthin}, or~\ref{gbthin}, is not justified.

Combining Equations~\ref{sbeq} and~\ref{intgb} one finds
\begin{equation}
T_\mathrm{int}=\frac{h\,\nu_0}{k_B}\,\left(\frac{h^3\,c^2}{2\,\pi\,\sigma\, k_B^4\,\zeta(4+\beta)\,\Gamma(4+\beta)}\right)^\frac{1}{\beta}\,.
\end{equation}
The above relation indicates that, if $\nu_0$ decreases (i.e. the greybody gets optically thin 
over a shorter range of frequencies), $T_\mathrm{int}$ decreases linearly as well, shortening
the range of temperatures $T < T_\mathrm{int}$ over which $W_\mathrm{b}(T)>W_\mathrm{g}(T)$,
i.e. the physically meaningful case.

This issue is originated by the fact that the integral that leads to Equation~(\ref{intgb}) 
is computed over all the 
frequencies, also those such that $\nu>\nu_0$, violating the optically thin assumption. 
Therefore this equation is still correct only if $\nu_0 \gg \nu_*$, where $\nu_*$ is 
the frequency such that
\begin{equation}\label{partialint}
\int_0^\infty\frac{x^{3+\beta}}{\mathrm{e}^{x}-1}\mathrm{d}x \approx
\int_0^{x_*}\frac{x^{3+\beta}}{\mathrm{e}^{x}-1}\mathrm{d}x\,
\end{equation}
where $x_*\equiv h\,\nu_*/k_B\,T$.

We postpone the discussion of this issue to Section~\ref{byproduct}, after having
approached in Section~\ref{lsubmm} the class of integrals like the one at the right 
hand side of the above equation. 

Finally, we numerically calculated the $W_{\mathrm{g}}(T)$ curves in the optically 
thick case (for which it is not possible to obtain an analytic relation), and show them in 
Figure~\ref{lbols}. Such curves $i$) do not show a power-law behavior as in the optically
thin case; $ii$) do not suffer of the issue of getting higher than the 
blackbody, being Equation~\ref{igb} valid at all frequencies; $iii$) are always smaller than 
the corresponding optically thin case; $iv$) at increasing $\nu_0$ and for opportunely 
low values of $T$ (as those probed in the figure), the optically thick and thin cases get 
practically indistinguishable (since the Equation~\ref{igbthin} is valid over most
of the frequency range).

\subsection{The $L_{\mathrm{bol}}/M$ ratio}
The ratio between the bolometric luminosity of a core/clump, due to the contribution
of a possible contained young stellar object and by the residual matter in the parental 
core/clump, and its mass $M$ is a largely used tool for characterizing the star formation ongoing 
in a such structure \citep{mol08,eli13}. Indeed, an increase of $L_{\mathrm{bol}}$ is
expected as the central source evolves and its temperature increases (so the emission peak
shifts towards shorter wavelengths); this is evident especialy during the main accretion phase
\citep[][and references therein]{mol08}. Dividing $L_{\mathrm{bol}}$ by the total envelope mass 
removes any dependence on the total amount of emitting matter. Interestingly, a 
$L_{\mathrm{bol}}/M$ built in this way is also a distance-independent quantity. It is 
important to show the relation between 
$L_{\mathrm{bol}}/M$ and $T$ for an optically thin greybody, which corresponds to the case
of a starless core/clump, to evaluate departures from this behavior, typical of proto-stellar
source.

On one hand, the bolometric luminosity of a greybody located at a distance $d$ is 
observationally evaluated starting from the measured flux:
\begin{equation}\label{bollum}
L_{\mathrm{bol}}=4\,\pi\,d^2 \int_0^\infty  F_{\nu}\,\mathrm{d}\nu\,.
\end{equation}
On the other hand, $L_{\mathrm{bol}}=W_{\mathrm{gb}}$, so using Equation~\ref{gbthin}
for $F_\nu$, one obtains

\begin{equation}\label{lmratio}
\begin{split}
&\frac{L_{\mathrm{bol}}}{M}&=&\frac{4\,\pi\,\kappa_{\mathrm{ref}}}{\nu_{\mathrm{ref}}^\beta}
\int_0^\infty \nu^\beta B_\nu(T)\,\mathrm{d}\nu=\\
& & = &4\,\kappa_{\mathrm{ref}} \left(\frac{\nu_0}{\nu_{\mathrm{ref}}}\right)^\beta\,W_{\mathrm{g}}=\\
& &=&\frac{8 \,\pi\, k_B^{4+\beta}\, \zeta(4+\beta)\,\Gamma(4+\beta)\,\kappa_{\mathrm{ref}}}{h^{3+\beta}\,c^2\,\nu_{\mathrm{ref}}^{\beta}}~T^{4+\beta}\,,
\end{split}
\end{equation}
which is dependent again on $T^{4+\beta}$, but independent on $\nu_0$, in the
limit of Equation~\ref{partialint}.

\subsection{The $L_{\mathrm{smm}}/L_{\mathrm{bol}}$ ratio}\label{lsubmm}
Another quantity involving the bolometric luminosity and used to characterize the evolutionary 
state of young stellar objects is the $L_{\mathrm{smm}}/L_{\mathrm{bol}}$ ratio \citep{and00},
where $L_{\mathrm{smm}}$ is the fraction of $L_{\mathrm{bol}}$ for the sub-millimeter domain,
i.e. for $\lambda$ larger than a certain $\lambda_{\mathrm{smm}}$. For example, with respect 
to the Class~0/I/II/III classification of low-mass young stellar objects \citep{lad84,lad87,and93}, \citet{and00} recognized as Class~0 those objects with $L_{\mathrm{smm}}/L_{\mathrm{bol}}> 0.005$,
for $\lambda_{\mathrm{smm}}= 350\mu$m. 

For an optically thin greybody, the dependence of this ratio on the greybody parameters
can be ascertained starting from Equation~\ref{bollum} as follows :
\begin{equation}\label{lratio}
\begin{split}
L_{\mathrm{smm}}/L_{\mathrm{bol}}&=&\frac{\int_0^{\nu_{\mathrm{smm}}} F_{\nu}\,\mathrm{d}\nu}{\int_0^\infty  F_{\nu}\,\mathrm{d}\nu}=\frac{\int_0^{x_{\mathrm{smm}}} \frac{x^{3+\beta}}{\left(\mathrm{e}^x-1\right)}\,\mathrm{d}x}{\int_0^\infty 
\frac{x^{3+\beta}}{\left(\mathrm{e}^x-1\right)}\,\mathrm{d}x}=\\
&=&\frac{\sum_{n=1}^\infty \frac{1}{n^{4+\beta}}\int_0^{nx_{\mathrm{smm}}} y^{3+\beta}\mathrm{e}^{-y}~\mathrm{d}y}
{\zeta(4+\beta) \Gamma(4+\beta)}\,,
\end{split}
\end{equation}
where $\nu_{\mathrm{smm}}=c/\lambda_{\mathrm{smm}}$ is the frequency assumed as the upper end of 
the sub-mm domain and $x_{\mathrm{smm}}\equiv h\,\nu_{\mathrm{smm}}/k_B T $. While the denominator of the 
last member does not depend on $T$ (see Equation~\ref{gammazeta}),  the 
numerator contains an integral with a finite upper integration limit containing in turn
the temperature, which requires a more complex treatment. One needs to invoke the concept 
of lower incomplete gamma function, defined as 
$\gamma(s,a)\equiv \int_0^a y^{s-1}\,\mathrm{e}^{-y}\,dy$. It is found \citep[e.g.,][]{pre07} that
\begin{equation}\label{gammasa}
\gamma(s,a)=a^s \mathrm{e}^{-a}{s}\,\sum_{i=0}^\infty\frac{\Gamma(s)}{\Gamma(s+1+i)}a^i\,,
\end{equation} 

Therefore, being in this case $s=\beta+4$ and $a=n\,x_{\mathrm{smm}}$,
\begin{equation}\label{inty}
\begin{split}
&\int_0^{nx_{\mathrm{smm}}} y^{3+\beta}\mathrm{e}^{-y}~dy=\\
&=\Gamma(4+\beta)(n\,x_{\mathrm{smm}})^{4+\beta}\,\mathrm{e}^{-n\,x_{\mathrm{smm}}}\,
\sum_{i=0}^{\infty}\frac{(n\,x_{\mathrm{smm}})^i}{\Gamma(5+\beta+i)}\,.
\end{split}
\end{equation}

So, Equation~\ref{lratio} becomes
\begin{equation}\label{lratiofin}
\begin{split}
L_{\mathrm{smm}}/L_{\mathrm{bol}}=
\frac{1 }{\zeta(4+\beta)}\,\sum_{n=1}^\infty \mathrm{e}^{-n\,x_{\mathrm{smm}}}\,
\sum_{i=0}^{\infty}\frac{n^i}{\Gamma(5+\beta+i)}x_{\mathrm{smm}}^{4+\beta+i}
\end{split}
\end{equation}

\begin{figure}
\includegraphics[width=8.3cm]{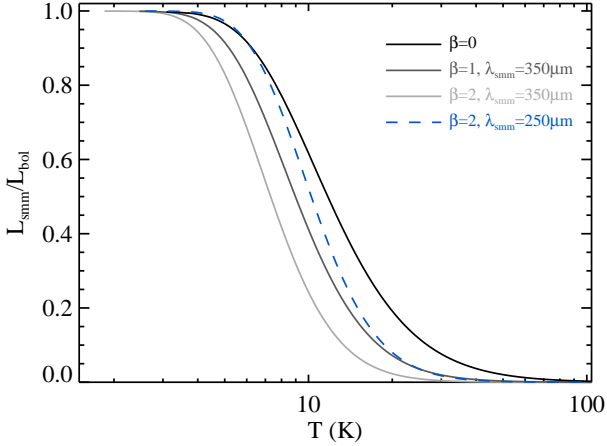}
\caption{Plot of the $L_{\mathrm{smm}}/L_{\mathrm{bol}}$ ratio vs $T$ relation, 
as expressed by Equation~\ref{lratiofin}. The grey lines correspond to 
$\lambda_{\mathrm{smm}}=350~\mu$m and different values of $\beta$: 0 (black), 
1 (dark grey), and 2 (light grey). The blue dashed line corresponds to the 
case $\lambda_{\mathrm{smm}}=250~\mu$m and $\beta=2$.}\label{tlsubmm}
\end{figure}

In Figure~\ref{tlsubmm} the behavior of $L_{\mathrm{smm}}/L_{\mathrm{bol}}$ vs $T$ is 
shown for different choices of $\beta$ and $\nu_{\mathrm{smm}}$.

\subsubsection{A by-product: discussing Equation~\ref{partialint}}\label{byproduct}
Here we exploit the results found for the integration of the greybody over a non-infinite range
(i.e., Equation~\ref{gammasa}) to conclude the discussion about Equation~\ref{partialint}, 
namely regarding the frequency $\nu_*$ such that, given $x_*=h\,\nu_1 / k_B\,T$, the condition, 
say, $R\equiv \int_0^{x_*} \frac{x^{3+\beta}}{\mathrm{e}^x-1}\,\mathrm{d}x / \int_0^{\infty} 
\frac{x^{3+\beta}}{\mathrm{e}^x-1}\,\mathrm{d}x> 99\%$ is satisfied. The ratio $R$ 
coincides exactly with the case of Equation~\ref{lratio} fully developed through 
Equation~\ref{lratiofin}, with $\nu_1$ playing in this case the role of $\nu_\mathrm{smm}$
in those equations.
 
For a precision of 10\% (i.e. $R=0.1$), $x_* \ge 8$ when $\beta=1$, and $x_* \ge 11$ for 
$\beta=3$. Turning $x_*$ in a wavelength, one finds $\lambda_*$($\mu$mm)$=1308/T$(K) and 
$\lambda_*$($\mu$m)$=1798/T$(K), for $\beta=3$ and 1, respectively. For instance, for a SED 
with $T=20$~K and $\beta=1$, the greybody must be already optically thin at 
$\lambda_*=90~\mu$m, while for $T=30$~K the limit on being optically thin 
is pushed down to $60~\mu$m.

Clearly, these values for $\lambda_*$ apply to SED known over the infinite range 
$\lambda_* \le \lambda < \infty$. To be less generic, let us consider the practical 
case of a SED which is known only at five \emph{Herschel} bands: the two 70~$\mu$m and 160~$\mu$m 
for PACS, and the three SPIRE bands 250~$\mu$m, 350~$\mu$m and 500~$\mu$m. This is the 
case for the \emph{Herschel} surveys already mentioned: GBS, HOBYS, and HIGAL. We derived the 
theoretical SEDs from Equation~\ref{igb} for $\lambda_0=10$, 50, and 100~$\mu$m, and for 
$\beta=1$,2 and 3; for the temperature we explored the range $5 \le T \le 50$~K 
in steps of 1~K. 

\begin{figure}
\includegraphics[width=8.0cm]{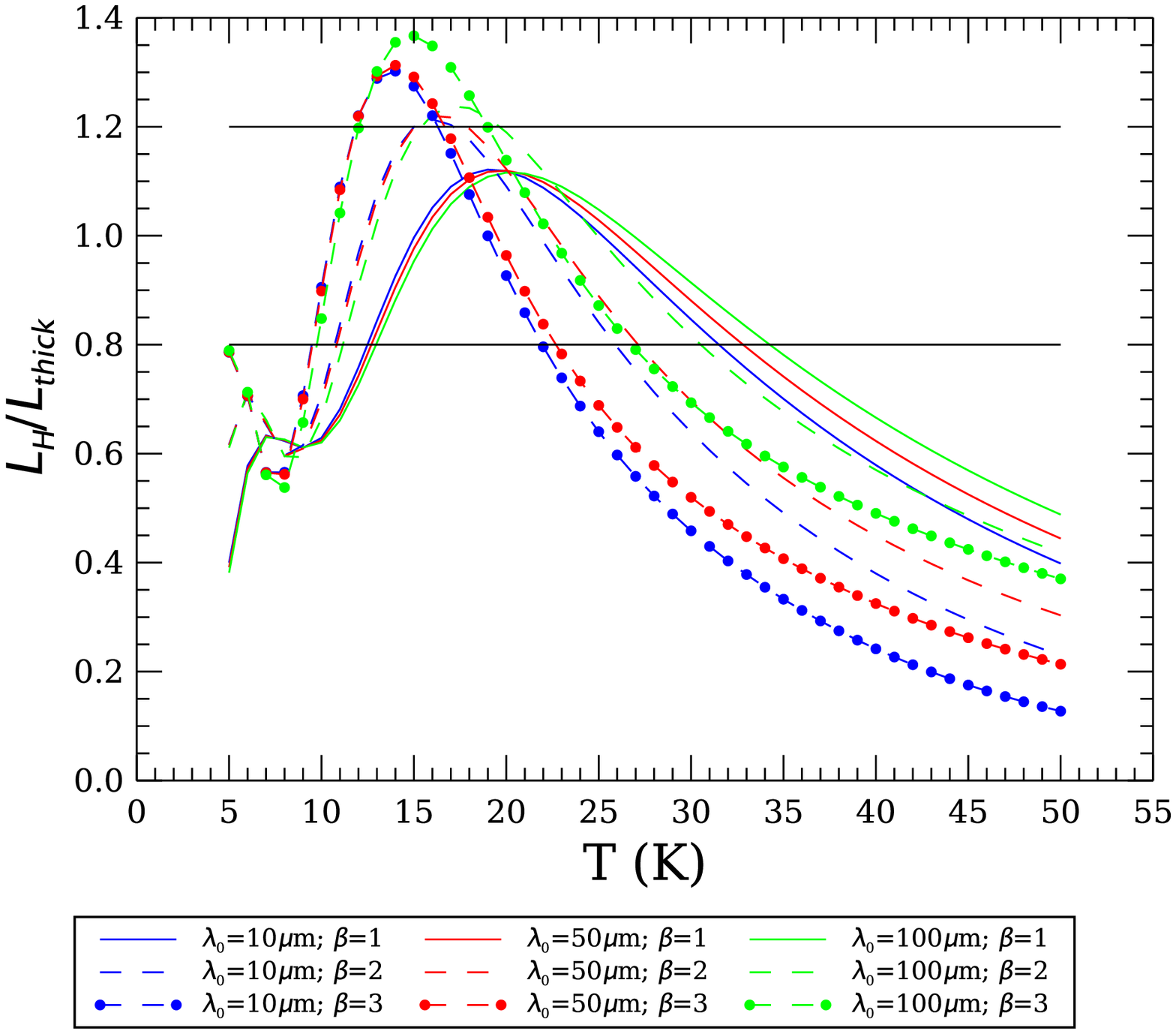}
\includegraphics[width=8.0cm]{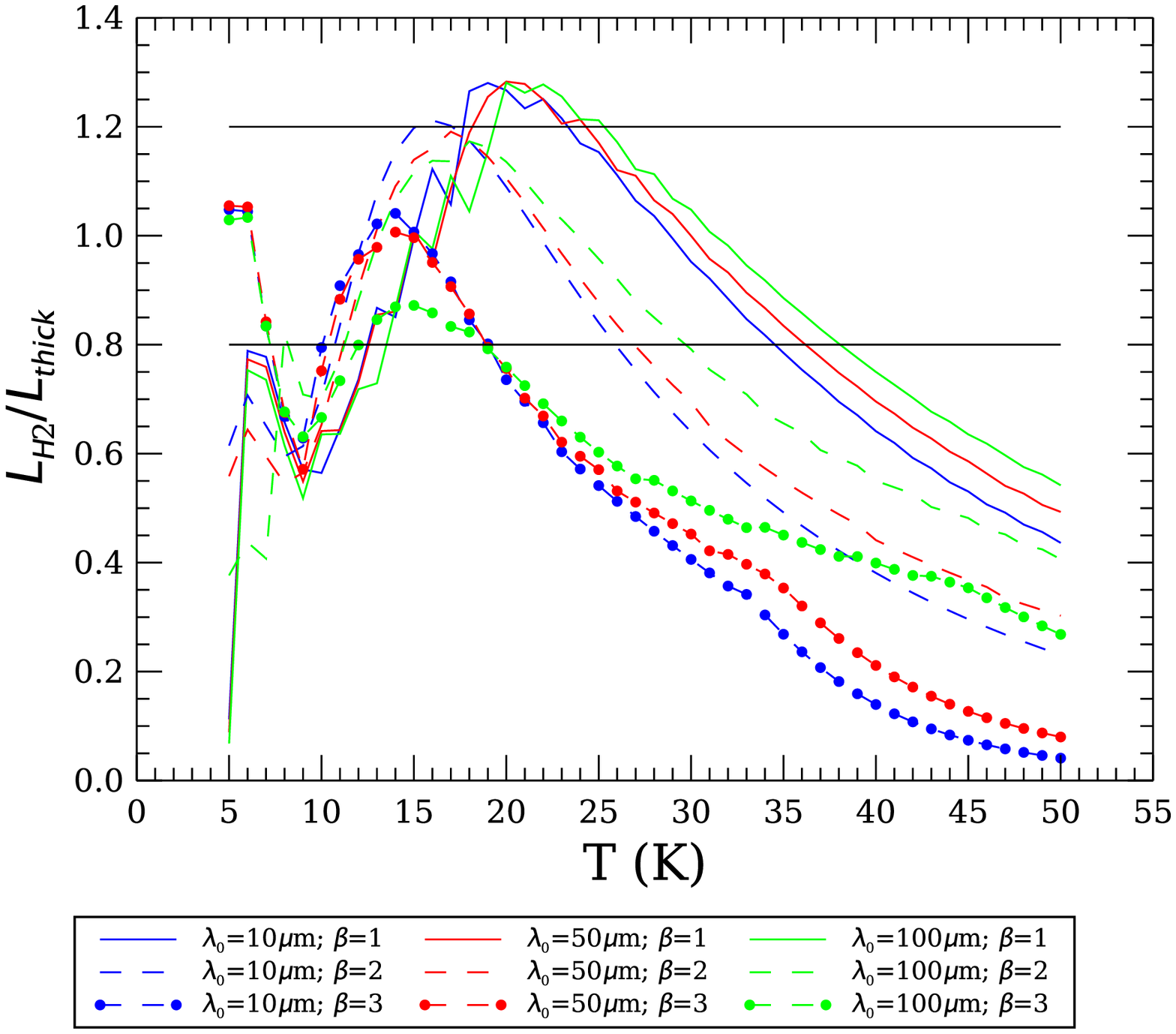}
\includegraphics[width=8.0cm]{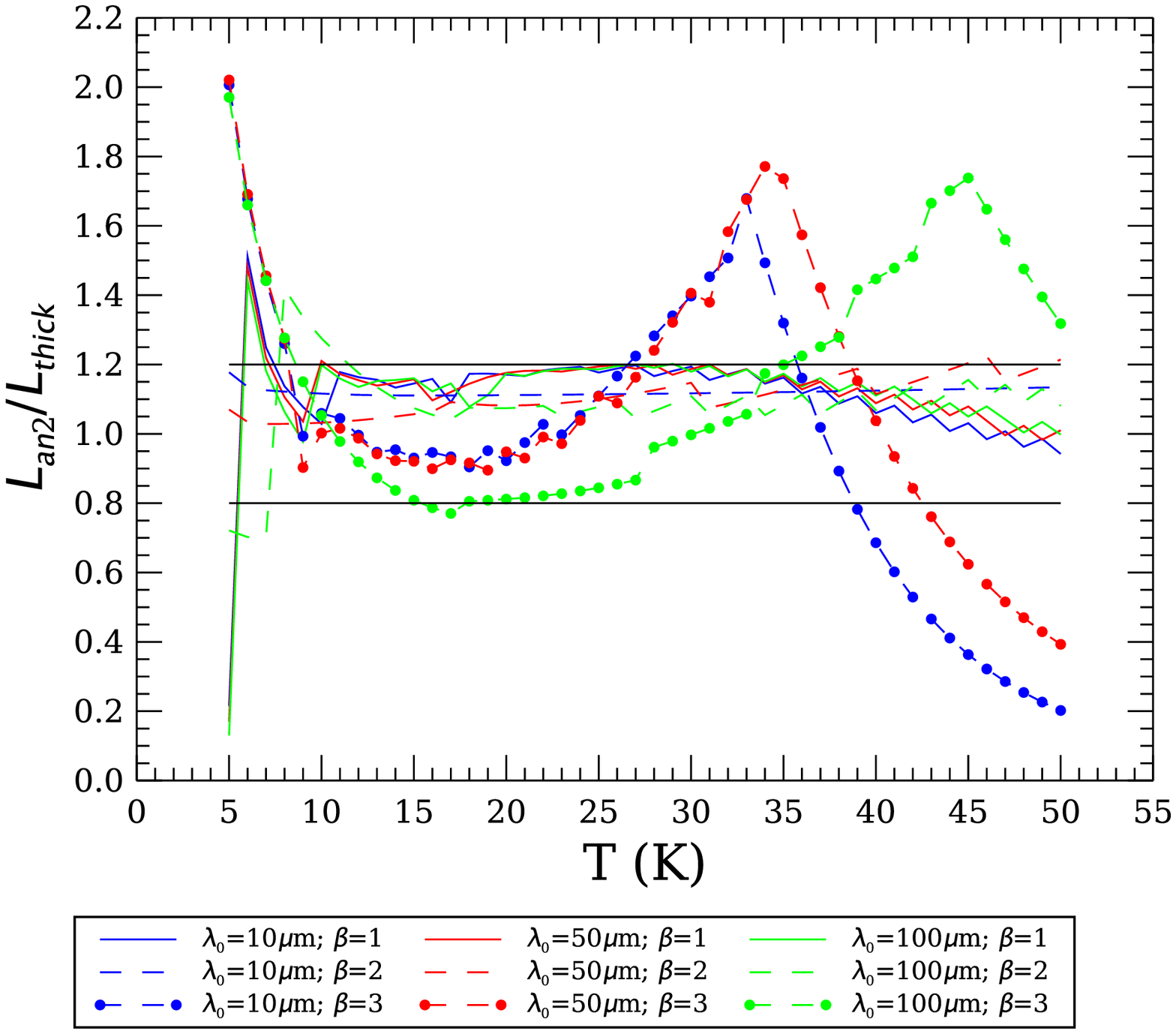}
\caption{Ratios between three different estimates of the luminosity of a SED and
$L_\mathrm{thick}$, i.e. the luminosity of a greybody obtained by numerical integration 
of Equation~\ref{igb} (considered here as the \textit{true} 
luminosity of the SED), vs greybody temperature. \textit{Top}: ratio between $L_\mathrm{H}$, 
obtained by integrating the SED observed by Herschel, and $L_\mathrm{thick}$. 
\textit{Centre}: ratio between $L_\mathrm{H2}$, obtained by numerical integration 
of an optically-thin model with $\beta=2$, best-fitting the SED observed by Herschel, 
and $L_\mathrm{thick}$. \textit{Bottom}: ratio between $L_\mathrm{an2}$, obtained directly 
through Equation~\ref{intgb} with $\beta=2$ and $T$ derived by the same fit as in the 
previous panel, and $L_\mathrm{thick}$. The horizontal black lines show the 
20\% agreement zone.}\label{luminosities}
\end{figure}

For each SED we computed the \textit{true} luminosity ($L_\mathrm{thick}$) by 
numerical integration of Equation~\ref{igb} from $1~\mu$m to 1~mm: in the upper panel 
of Figure~\ref{luminosities} we show the ratio between $L_\mathrm{thick}$ and the 
luminosity $L_\mathrm{H}$ computed integrating the five-band Herschel SED. 
This figure shows the error\footnote{We stress that this is an error and not 
an uncertainty.} associated to a Herschel-derived luminosity. However, this 
is only of mathematical interest because in the most common case the astronomer 
does not know $\beta$, so that it is not known with which curve $L_\mathrm{H}$ 
should be compared.

The other two panels are more interesting because we compared the true 
luminosity with two quantities derivable from the data. Since we are assuming 
that only the five Herschel fluxes are known (we used five bands, but including 
the 100~$\mu$m PACS band too would not alter our conclusions), it is not possible 
to derive a robust estimate of $\beta$ directly from the observed values, so 
that we fix $\beta=2$, a common choice when dealing only with Herschel data
\citep{sad13}. For any theoretical SED, we looked for the best-fitting optically 
thin greybody with $\beta=2$. For this greybody, we computed both the luminosity 
$L_\mathrm{H2}$ obtained integrating only the fluxes at the five considered 
wavelengths, and the luminosity $L_\mathrm{an2}$ given by Equation~\ref{intgb}. 
The ratios between these two luminosities and the true luminosity $L_\mathrm{thick}$ 
are shown in the central and bottom panels of Figure~\ref{luminosities}, 
respectively. For simplicity, the $x$-axis reports the true temperature in 
both cases, although the temperature used to compute the luminosity is that 
derived from the fit. The central panel of Figure~\ref{luminosities} is, qualitatively, 
quite similar to the top panel and shows an erratic behaviour of the ratio: the agreement 
is within 20\% (limit shown by means of the two black horizontal lines) for $T \ga 10$~K, 
but the upper limit on $T$, where the agreement is good, strongly depends on $\beta$, 
which is unknown. The bottom panel, on the contrary, shows a ratio 
contained in the 20\% limits for all the $T \ga 10$~K up to 50~K, the highest 
$T$ used in the synthetic SEDs. Only for $\beta$ as high as 3 there are ranges 
of $T$ for which the agreement is not good.

Our conclusion is that once an astronomer decides to fit an observed SED 
with an optically thin greybody with $\beta=2$, it is better to compute 
the luminosity from Equation~\ref{intgb} rather than to integrate the 
observed fluxes, or those derived from the fit. The case in which 
$\beta$ is known from the data is dealt with in Section~\ref{sedbetatemp}.

\section{Bolometric temperature}\label{tbolsec}
The bolometric temperature $T_{\mathrm{bol}}$ \citep{mye93} is another quantity used in 
the study of star formation to quantify the evolutionary status of young stellar objects. 
Indeed it constitutes an estimate, in units of temperature, of the ``average frequency'' 
of a SED (with the fluxes composing the SED used as weights):
\begin{equation}
\bar{\nu}=\frac{\int_0^\infty \nu\,I_{\nu}~\mathrm{d}\nu}
{\int_0^\infty~I_{\nu}~\mathrm{d}\nu}\,.
\end{equation}

For a blackbody, exploiting Equation~\ref{intgb}, one finds
\begin{equation}\label{numed}
\bar{\nu}_{\mathrm{bb}}=\frac{\int_0^\infty \nu\,B_{\nu}(T)~\mathrm{d}\nu}
{\int_0^\infty~B_{\nu}(T)~\mathrm{d}\nu}=
\frac{4\,k_B}{h}\frac{\zeta(5)}{\zeta(4)}\,T\,.
\end{equation}

Rigorously, the bolometric temperature of a generic source is defined as the temperature 
of a blackbody having the same mean frequency $\bar{\nu}$:
\begin{equation}\label{tboleq}
T_{\mathrm{bol}}=\frac{h}{4\,k_B}\,\frac{\zeta(4)}{\zeta(5)}\bar{\nu}=
\frac{h}{4\,k_B}\,\frac{\zeta(4)}{\zeta(5)}\,
\frac{\int_0^\infty \nu\,F_{\nu}~\mathrm{d}\nu}{\int_0^\infty~F_{\nu}~\mathrm{d}\nu}\,.
\end{equation}

In this definition, \citep{mye93} adopted a normalization suggested by Equation~\ref{numed}
to obtain $T_{\mathrm{bol}}=T$ for the blackbody case.

Looking at different phases of star formation, the transition from Class~0 
to Class~I and then to Class~II sources is characterized by a temperature getting 
warmer and warmer and the SED getting brighter and brighter in the near- and mid-infrared: 
as a consequence of this, also $\bar{\nu}$ and $T_{\mathrm{bol}}$ increase. In this 
respect, \citet{che95}, suggested to identify the aforementioned evolutionary classes  
through $T_{\mathrm{bol}}$.

\begin{figure}
\includegraphics[width=8.3cm]{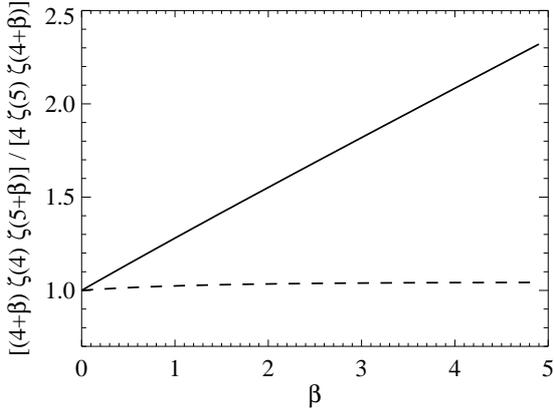}
\caption{Solid line: behavior of the multiplicative term in front of $T$ in the expression 
of $T_{\mathrm{bol}}$ (Equation~\ref{boltemp}), namely 
$[(4+\beta)\zeta(4)\zeta(5+\beta)]/[4\,\zeta(5)\zeta(4+\beta)]$, as a function of $\beta$. 
Dashed line: behavior of 
$\zeta(5+\beta) / \zeta(4+\beta)$  as a function of $\beta$.}\label{tbolfactor}
\end{figure}

Here we explore the analytic behavior of the bolometric temperature of an optically thin 
greybody as a function of the various involved parameters. Combining 
Equations~\ref{numed}, \ref{gbthin}, and~\ref{intgb}, one obtains
\begin{equation}\label{numedgb}
\begin{split}
\nu_{\mathrm{gb}}&= \frac{\int_0^\infty \nu^{1+\beta}~B_{\nu}(T)~\mathrm{d}\nu}
{\int_0^\infty~\nu^{\beta}~B_{\nu}(T)~\mathrm{d}\nu}=&\\
&= \frac{k_B \zeta(5+\beta) \Gamma(5+\beta)}{h\,\zeta(4+\beta) \Gamma(4+\beta)}\,T=&\\
&= \frac{k_B}{h}\frac{\zeta(5+\beta)}{\zeta(4+\beta)}\,T\,,&
\end{split}
\end{equation}
so that the bolometric temperature for a greybody is given by
\begin{equation}\label{boltemp}
T_{\mathrm{bol}}= \frac{4+\beta}{4}\frac{\zeta(4)\zeta(5+\beta)}{\zeta(5)\zeta(4+\beta)}\,T\,.
\end{equation} 
It is noteworthy that in this relation the proportionality factor between $T_{\mathrm{bol}}$
and $T$ is a monotonically increasing function of $\beta$ being in turn the product of two 
increasing functions, as illustrated in Figure~\ref{tbolfactor}.

Furthermore, since $\zeta(4)/\zeta(5)\simeq 1.044$, for $\beta=1$ the ratio of the 
$\zeta$ functions differ from 1 by $\sim$2\%, and by $\sim$1\% for $\beta=2$. One makes 
then a very small error putting 
$\zeta(5+\beta)/\zeta(4+\beta)=1$, so that
\begin{equation}\label{TforG}
T_\mathrm{bol}=\frac{4+\beta}{4}\,T ~\,\,(\mathrm{for\ }\beta\ge1)\,.
\end{equation}
As an immediate consequence, a greybody source with $\beta=2$ and $T>47$~K has 
$T_\mathrm{bol}>70$~K, i.e. above the boundary between Class~0 and Class~I established 
by \citet{che95}.

\section{An alternative way to estimate $T$ and $\beta$ for an observed SED}\label{sedbetatemp}
Equation~\ref{numedgb} represents the the first moment of the distribution of 
an optically thin greybody, hereafter $\nu_1$. In general,
the $n$-th moment can be straightforwardly calculated as:
\begin{equation}\label{nmoment}
\nu_n=\left(\frac{k_B\,T}{h}\right)^n\frac{\Gamma(n+4+\beta)\zeta(n+4+\beta)}{\Gamma(4+\beta)\zeta(4+\beta)}\,.
\end{equation}
So, computing now the second moment one finds
\begin{equation}
\begin{split}
\nu_2&=\left(\frac{k_B\,T}h\right)^2\frac{\Gamma(6+\beta)\zeta(6+\beta)}
{\Gamma(4+\beta)\zeta(4+\beta)}\,.
\end{split}
\end{equation}
As shown above, one can put $\zeta(6+\beta)/\zeta(4+\beta)\sim1$ so that
\begin{equation}\label{nu2}
\nu_2=(5+\beta)(4+\beta)\left(\frac{k_B\,T}h\right)^2 \,\,\,\,\,(\mathrm{for\ }\beta\ge1)\,.
\end{equation}
Combining Equations~\ref{numedgb} and \ref{nu2}, one finds the interesting relation
\begin{equation}\label{bforG}
\beta=\frac{5\nu_1^2-4\nu_2}{\nu_2-\nu_1^2}\,.
\end{equation}

In the same way, by a combination of these two frequencies (actually $\nu_2$ has the unit 
of Hz$^2$) it is possible to derive the formula for the temperature
\begin{equation}\label{tnu2}
T=\frac{h}{k_B}\left(\frac{\nu_2}{\nu_1}-\nu_1\right)\,.
\end{equation}

These two equations give the exact values of $\beta$ and $T$, provided that the spectrum 
is known over a wide range of frequencies (or wavelengths). In reality this is not always 
true: the smaller the number of data points, the higher the error associated with these 
equations.

To give an idea of the applicability of these two formulae, we took from 
literature the case of the the candidate first-hydrostatic core CB17MMS \citep{che12}: 
the authors report the fluxes at 100~$\mu$m, 160~$\mu$m, 850~$\mu$m and 1.3~mm, so 
just four wavelengths. From these data we computed, with a straightforward application 
of the trapezium rule, $\nu_1$ and $\nu_2$. As a second step, we generated a grid of 
models, with $5 \le T(\mathrm{K}) \le 50$ and $1 \le \beta \le 3$, at the same four 
aforementioned wavelengths. For each model we computed the expected values 
$\bar{\nu_1}$ and $\bar{\nu_2}$, that are compared with the values derived from the 
observations $\nu_1$ and $\nu_2$: we looked for the minimum of the residuals defined as
\begin{equation}
\delta\equiv\left(\frac{(\bar{\nu_1}-\nu_1)}{\mathrm{max}(\bar{\nu_1})}\right)^2+
\left(\frac{(\bar{\nu_2}-\nu_2)}{\mathrm{max}(\bar{\nu_2})}\right)^2\,,
\end{equation}
where the normalization is necessary due to the fact that, for the same model, 
$\bar{\nu_2}$ is of the order of $\bar{\nu_1}^2$: indeed, without this normalization,
$\delta$ would be dominated by the term containing $\nu_1$. All the models in the grid, 
constructed in steps of $0.1$~K in $T$ and $0.1$ in $\beta$, were sorted by 
increasing residuals. The best model, corresponding to the lowest $\delta$,
provides $T$ and $\beta$, whereas the ten best models are taken into account to
evaluate the spread of these two parameters, hence the uncertainty affecting them.
The final result is $T=10.6 \pm 1.3$~K and 
$\beta=2.1\pm0.6$. These values are in good agreement with those reported by \citet{che12}, 
namely $T\sim10$~K and $\beta=1.8$. Notice that if we would have just used $\nu_1$ and 
$\nu_2$ to derive directly $T$ and $\beta$, the result would have been $T=7.8$~K 
and $\beta=6.5$, which is unphysical.

Now, if we measure the luminosity of the object, again through the trapezium rule, 
we find $L=0.18~L_{\odot}$, at the distance of 250~pc. This luminosity can be compared 
with that expected theoretically (Equation~\ref{intgb}) which depends on $T$, $\beta$, 
the solid angle and $\nu_0$; unfortunately \citet{che12} did not provide the solid angle 
in their paper, but the fluxes they reported for CB17MMS were derived with apertures 
ranging from 10\arcsec to 20\arcsec: estimating the solid angle from these apertures, 
Equation~\ref{intgb} tells us that the wavelength at wich $\tau=1$ is in the range
$30 \lesssim \lambda_0(\mu\mathrm{m}) \lesssim 58$, which looks reasonable.

As another example we consider the SED of GG~Tau~A as reported by \citet{sca13}: 
in this case the SED is emitted by a disc so that the hypothesis of a single-temperature 
optically-thin greybody is very coarse, but still our results can be compared with 
those of the author. Our procedure, applied to the SED from 100~$\mu$m to 1.86~cm, 
gives $T=18.5 \pm 6.7$~K and $\beta=1.1 \pm 1.7$. Our $T$ agrees well with the reported 
value of $T=19.42 \pm 0.55$~K; the value of $\beta$ has a large uncertainty but the 
best-fit value, $\beta=1.1$, is also in agreement with the value of \citet{sca13}, $0.96\pm0.04$. 
Since $T(R=300\mathrm{AU}) \approx 20$~K, as reported by the author, we took this radius 
as an estimate of the solid angle, given the distance of 140~pc, and we found 
$\lambda_0\sim29$~$\mu$m.

As a last example, we used the compilation of fluxes reported recently by 
\citet{ren16} for a set of sources in NGC~2024 whose SEDs are built from 70  
to 850~$\mu$m. In Table~\ref{intT} we reported the name of each source, 
$T$, $\beta$ and $L_\mathrm{bol}$ as computed by the authors, the same quantities as 
computed by us, and, in the last column, $\lambda_0$.

The source FIR-1 gives a result compatible with the hypothesis of an 
optically thin greybody ($\lambda_0 < \lambda_{\mathrm{min}}$) only if 
set $\lambda_\mathrm{min}=160$~$\mu$m, i.e., after discarding the first two 
available wavelengths; the same happens with FIR-2 and FIR-3 as well. 
For FIR-4 and FIR-6 the whole SED has been used, as the resulting 
$\lambda_0$ is smaller than 70~$\mu$m. FIR-5 is resolved in two sources at 
450~$\mu$m and not resolved at the other wavelengths: for this reason we
decided not to consider this source. Finally, the SED of FIR-7 has 
been limited to $\lambda \ge 250~\mu$m to fulfill the condition 
$\lambda_0 < \lambda_\mathrm{min}$.


Clearly, our $L_\mathrm{bol}$ are smaller than values of \citet{ren16}  because they 
were evaluated over a shorter range of wavelengths: if we trust the values of 
$\lambda_0$ found, our luminosities constitute an estimate of the optically-thin 
contribution to $L_\mathrm{bol}$ for each source. The large uncertainty in 
$T$ for FIR-7 clearly reflects the fact that we derived the physical parameters 
of this source at large wavelengths, excluding the peak of the SED.

In the examples reported above we have shown in a number of cases how well, 
or how bad, our Equations 46 and 47 can be used to extract physical informations from 
a SED: one may wonder why
Equations~\ref{bforG} and~\ref{tnu2} should be used to find $\beta$ 
and $T$ instead of using well-known routines that can solve the non-linear least-squares 
problems. There are a few advantages, indeed: first, one does not need to give initial 
values for the parameters, which not always are obvious to be estimated. Second, specifically 
to the greybody problem, it is often assumed that, for \emph{Herschel} data, it is not possible to 
have realistic estimates of both $\beta$ and $T$, given the well-known degeneracy between these 
two values \citep[][]{juv13}, as it can be seen in Equation~\ref{numax}. On the contrary, 
our formulae do not imply a fitting procedure, and give the two parameters without being
affected by degeneracy. 
Third, if the distance and the solid angle are known for a source, one can derive also $\lambda_0$, 
i.e. the wavelength at which $\tau=1$. In the usual formalism given by, e.g., Equation~\ref{gbthin}, 
there is no way to derive $\lambda_0$ from the data. As a consequence of this, with our method the astronomer 
can judge a posteriori if the derived values are consistent with the optically-thin hypothesis, 
something that seldom is done in literature.

Of course, we should not forget that inferring $T$ and $\beta$ from real observations is 
more challenging because of line of sight mixing of temperature (breaking the condition of isothermal 
emission), asymmetric illumination of target source, and contribution of different population of 
dust grains to the net emission. But these caveats affect any kind of fitting procedure.

\begin{table*}
\caption{Results of applying Equations~\ref{bforG} and~\ref{tnu2} to source SEDs
of \citet{ren16}$^\mathrm{a}$.\label{intT}} 
\begin{tabular}{lcccccccc}
\hline
     &\multicolumn{3}{c}{\citet{ren16}}    &         & \multicolumn{4}{c}{This paper}\\
     \cline{1-4}\cline{6-9}
Name & $T$ & $\beta$ & $L_\mathrm{bol}$ & & $T$ & $\beta$ & $L_\mathrm{bol}$ & $\lambda_0$ \\
     & K   &         & $L_\odot$        & &  K  &         & $L_\odot$        & $\mu$m \\ 
\hline
FIR-1&18.5&2.5&80&&$29.3\pm8.7$&$1.6\pm0.4$&22&19\\
FIR-2&18.0&2.7&130&&$18.4\pm4.3$&$2.4\pm0.6$&32&145\\
FIR-3&17.5&2.6&220&&$19.0\pm3.4$&$2.3\pm0.7$&47&141\\
FIR-4&22.0&2.9&570&&$32.2\pm3.1$&$1.3\pm0.3$&221&64\\
FIR-6&18.5&2.7&160&&$26.1\pm2.4$&$1.0\pm0.4$&118&43\\
FIR-7&18.0&2.6&110&&$19\pm12$&$2.8\pm0.7$&8&70\\
\hline
\multicolumn{9}{l}\footnotesize{$^\mathrm{a}$ We used the whole SED, from 70~$\mu$m to 850~$\mu$m, for FIR-4 and 6;}\\ 
\multicolumn{9}{l}\footnotesize{from 160~$\mu$m to 850~$\mu$m for FIR-1, FIR-2, and FIR-3;}\\
\multicolumn{9}{l}\footnotesize{from 250~$\mu$m to 850~$\mu$m for FIR-7.}\\
\end{tabular}
\end{table*}

%

Finally, we provide an example of application of Equation~\ref{numed} to derive $T$ 
for a blackbody from an observed SED. We consider the COBE-FIRAS 
spectrum of the cosmic microwave background 
radiation measured by \citet{fix96}\footnote{Data are available at 
http://lambda.gsfc.nasa.gov/data/cobe/firas/monopole\_spec/firas\_monopole\_spec\_v1.txt.}. 
The temperature derived through Equation~\ref{numed} is 2.82~K, 
only 3\% higher than the 2.725~K value used by the authors to derive the monopole spectrum.

\section{Conclusions}\label{conclusions}
In this paper we collected and re-arranged a number of dispersed analytic relations
among the parameters of a greybody, developing further equations and discussing the errors 
involved by typical approximations. This is certainly of some interest for astronomers
who model the Galactic and extragalactic cold dust emission as a greybody, especially
in the current ``\emph{Herschel} era'', characterized by the availability of huge 
archives of photometric far-infrared data. In particular, 
\begin{itemize}
\item[-] The position of the peak of the greybody
emission, in terms of both frequency and wavelength as a function of the temperature
has been revised, considering deviations from the classical blackbody. Approximated 
expressions for it are suggested in correspondence of different regimes of optical 
thickness.

\item[-] Quantities typically exploited in the study of early phases of star formation
have been discussed in the case of an optically thin greybody. 
The bolometric luminosity of a greybody shows a power-law dependence on the temperature,
with exponent $4+\beta$, representing a general case of which the Stefan-Boltzmann's law 
is a particular case for $\beta=0$ (blackbody). 

\item[-] The ratio between the so-called sub-millimeter luminosity
and the bolometric one, which is used to recognize Class~0 young stellar objects, shows
a more complex behavior. The temperature at which this ratio gets larger than 0.05\% 
(so early-phase star forming cores/clumps are identified) decreases at increasing $\beta$. 

\item[-] The bolometric temperature of a greybody
is found to be linearly related to the temperature, through a multiplicative
constant that depends only on $\beta$ and can be further simplified for $\beta \geq 1$.  

\item[-] We indicate a method to derive the temperature and the dust emissivity
law exponent of a greybody, or simply the temperature of a blackbody, 
modeling an observed SED without performing a best-fit procedure.
We report and discuss the conditions for the applicability of this method, which appears
well suitable for well-sampled SEDs and in the range of temperatures typical of cold dense 
cores/clumps.

\end{itemize}

\section*{Acknowledgements}
We thank the anonymous referee for her/his accurate review and highly appreciated 
comments and suggestions, which significantly contributed to improving the quality of 
this paper.
D.E.'s research activity is supported by the VIALACTEA Project, a Collaborative Project 
under Framework Programme 7 of the European Union funded under Contract $\# 607380$, that 
is hereby acknowledged.


\bibliographystyle{mnras}
\bibliography{gb}


\bsp	
\label{lastpage}
\end{document}